\overfullrule=0pt
\input harvmac
\def\a{{\alpha}}

\def\ad{{\dot a}}
\def\bd{{\dot b}}
\def\cd{{\dot c}}
\def\l{{\lambda}}

\def\L{{\Lambda}}
\def\b{{\beta}}

\def\g{{\gamma}}

\def\d{{\delta}}

\def\s{{\sigma}}

\def\p{{\partial}}
\def\half{{1\over 2}}
\def\t{{\theta}}
\Title{ \vbox{\baselineskip12pt
\hbox{IFT-P.041/2001}}}
{\vbox{\centerline
{Lorentz Invariance of the Pure Spinor }
\centerline{BRST Cohomology for the Superstring}}}

\smallskip
\centerline{Nathan Berkovits\foot{e-mail: nberkovi@ift.unesp.br}}
\smallskip
\centerline{\it 
Instituto de F\'{\i}sica Te\'orica, Universidade Estadual Paulista}
\centerline{\it
Rua Pamplona 145, 01405-900, S\~ao Paulo, SP, Brasil}
\bigskip
\centerline{Osvaldo Chand\'{\i}a\foot{e-mail:
ochandia@maxwell.fis.puc.cl}}
\smallskip
\centerline{\it Facultad de F\'{\i}sica, Pontificia Universidad
Cat\'olica
de Chile}
\centerline{\it Casilla 306, Santiago 22, Chile}

\bigskip

\noindent

In a previous paper, the BRST cohomology in the pure spinor
formalism of the superstring was shown to coincide with the light-cone
Green-Schwarz spectrum by using an $SO(8)$ parameterization of the
pure spinor. In this paper, the $SO(9,1)$ Lorentz generators are
explicitly constructed using this $SO(8)$ parameterization, proving
the Lorentz invariance of the pure spinor BRST cohomology.

\Date{May 2001}

\newsec{Introduction}

Recently, the superstring was covariantly quantized using the
BRST operator $Q=\int \l^\a d_\a$ where $d_\a$ is the
fermionic Green-Schwarz constraint and $\l^\a$ is a pure
spinor satisfying
\eqn\pure{\l^\a\g^\mu_{\a\b}\l^\b=0} 
for $\mu=0$ to 9 \ref\berk
{N. Berkovits, ``Super-Poincar\'e Covariant Quantization of the
Superstring,'' JHEP 04 (2000) 018, hep-th/0001035}.
In order to prove equivalence of the cohomology of $Q$
with the light-cone Green-Schwarz spectrum, it was useful to solve
the pure spinor constraint of \pure\ using an $SO(8)$ parameterization
of $\l^\a$ and rewrite $Q$ in terms of unconstrained 
variables \ref\cohom{N. Berkovits, ``Cohomology in the Pure Spinor 
Formalism for the Superstring,'' JHEP 09 (2000) 046, hep-th/0006003.}. This
$SO(8)$ parameterization of $\l^\a$ is more complicated than the 
$U(5)$ parameterization of \berk\ since it involves an infinite
number of gauge degrees of freedom. However, it was necessary for the
cohomology computation since the $U(5)$ parameterization becomes singular
at certain values of $\l^\a$.

In this paper, $SO(9,1)$ Lorentz generators will be explicitly constructed
out of these unconstrained $SO(8)$ variables, thereby proving Lorentz
invariance of the cohomology computation. Although part of this
construction already appeared in \cohom, the most complicated Lorentz
generator, $M^{j-}$, was left incomplete. As will be shown here, verifying
that $M^{j-}$ satisfies $[M^{j-},M^{k-}]=[M^{j-},Q]=0$ involves several
rather impressive cancellations.

\newsec{$SO(8)$ Variables}

As discussed in \cohom, the pure spinor constraint of \pure\ for
$\l^\a$ can be solved in terms of $SO(8)$ variables $s^a$ and $v^j$
satisfying $s^a s^a=0$ as
\eqn\param{(\g^+\lambda)^a = s^a , \quad
(\g^-\lambda)^\ad = \s_j^{a\ad} v^j s^a,}
where $\g^\pm = \half(\g^0\pm\g^9)$, $\s_j^{a\ad}$ are $SO(8)$ Pauli
matrices satisfying $\s_{(j}^{a\ad}\s_{k)}^{b\ad}=2\d_{jk}\d^{ab}$,
and $(j,a,\dot a)=1$ to 8 are $SO(8)$ vector, chiral and anti-chiral
indices. The gauge invariance $\d v^j =\s^j_{a\dot a} s^a \epsilon^{\dot a}$
of the parameterization of \param\ leads to an infinite chain of
ghosts-for-ghosts $(s^a,v^j_M,t^\ad_M)$ for $M=0$ to $\infty$, 
and their conjugate
momenta $(r^a, w^j_M, u^\ad_M)$, where 
$v_0^j = v^j$ of \param,
$(s^a,r^a,v^j_M,w^j_M)$ are bosons and 
$(t^\ad_M,u^\ad_M)$ are fermions.
Also, the condition $s^a s^a=0$ can be treated as a BRST constraint
by introducing the fermionic ghost and anti-ghost $(b,c)$.

In terms of these unconstrained $SO(8)$ variables, it was
shown in \cohom\ that the BRST operator
$Q=\int \l^\a d_\a$ can be rewritten as 
\eqn\brst{Q' = \int  (s^a G^a - b s^a s^a + cT)}
where
\eqn\const{ G^a = (\g^- d)^a + \s_j^{a\ad} v^j_0 (\g^+ d)^\ad
+\widehat G^a,}
$$T= \half\Pi^- + v^j_0\Pi^j +\half
v^j_0 v^j_0 \Pi^+ + t^\ad_0 (\g^+ d)^\ad + \widehat T,$$
\eqn\ghat{\widehat G^a = \s_j^{a\ad} 
\sum_{M=0}^\infty (w^j_M t^\ad_M + v^j_{M+1} u^\ad_M),\quad
\widehat T=
\sum_{M=0}^\infty (v^j_{M+1} w^j_{M} + t^\ad_{M+1} u^\ad_{M}),}
and $d_\a$ and $\Pi^\mu$ are the fermionic and bosonic super-Poincar\'e
covariant momenta.
$SO(9,1)$ Lorentz generators will now be defined which commute with $Q'$,
proving the Lorentz invariance of the BRST cohomology.

\newsec{$SO(9,1)$ Lorentz Generators}

The $SO(9,1)$ Lorentz generators will be defined as 
\eqn\lorentz{M^{\mu\nu}=
\int (L^{\mu\nu} + N^{\mu\nu})}
where $L^{\mu\nu}= x^\mu \p x^\nu +
\half \t\g^{\mu\nu} p$ is constructed in the usual manner from the
$(x^\mu,\t^\a,p_\a)$ superspace variables and $N^{\mu\nu}$ is
constructed from the unconstrained $SO(8)$ variables of section 2.
It will now be shown that
\eqn\lore{N^{jk}=\half s^a (\s^{jk})_{ab} r^b +\sum_{M=0}^\infty
[v_{M}^{[j} w^{k]}_M +\half
t_{M}^\ad (\s^{jk})_{\ad\bd} u^\bd_{M}],}      
$$N^{j+}= w^j_{0},$$
$$N^{+-}=bc -\half s^a  r^a +\sum_{M=0}^\infty [ (M+1)
v_{M}^j w^j_{M} + (M+{3\over 2})
t_{M}^\ad w^\ad_{M}], $$
\eqn\lj{N^{j-}=-3\p v_{0}^j - v^k_{0} N^{jk}-
v^j_{0} N^{+-} -\half  v^k_{0} v^k_{0} w^j_{0}
+ v^j_{0} v^k_{0} w^k_{0} +\half c \s^j_{a\ad} t_{0}^\ad r^a }
$$+ \sum_{M,N=1}^\infty A^{MN j}_{klm} v^k_{M} v^l_{N} w^m_{M+N}
+ \sum_{M=1}^\infty \sum_{N=0}^\infty B^{MN j}_{k\ad\bd}
v^k_{M} t^\ad_{N} u^\bd_{M+N} + 
\sum_{M,N=0}^\infty C^{MN j}_{\ad\bd k}
t^\ad_{M} t^\bd_{N} w^k_{M+N +1}, $$
satisfy the $SO(9,1)$ current algebra 
\eqn\nope{N^{\mu\nu}(y) N^{\rho\sigma}(z) \to      
{{\eta^{\rho[\nu} N^{\mu]\s}(z) -
\eta^{\s[\nu} N^{\mu]\rho}(z) }\over {y-z}} - 3       
{{\eta^{\s[\mu} \eta^{\nu]\rho} }\over{(y-z)^2}}  }
where the constant $SO(8)$-covariant coefficients
$(A^{MN j}_{klm},
B^{MN j}_{k\ad\bd},
C^{MN j}_{\ad\bd k})$ will be determined in section 4 by requiring
that $[\int  M^{j-},Q']=0$.

To show that $N^{\mu\nu}$ satisfies \nope, the free-field OPE's 
\eqn\OPE{r^a(y) s^b(z) \to {{\d^{ab}}\over{y-z}},
\quad w^j_{M}(y) v^k_{N}(z)
\to {{\d^{jk} \d_{MN}}\over{y-z}},
\quad u^\ad_{M}(y) t^\bd_{N}(z)
\to {{\d^{\ad\bd} \d_{MN}}\over{y-z}},}
will be used. The only non-trivial part of checking the current algebra
involving $(N^{jk},N^{j+}, N^{+-})$ are the double poles of
$N^{jk}$ with $N^{jk}$ and $N^{+-}$ with $N^{+-}$.
The double pole of $N^{jk}$ with $N^{jk}$ gets a contribution of $+2$
from the first term and 
\eqn\regtwo{+2-2+2-2+... = 2\sum_{M=0}^\infty (-1)^M 
=2\lim_{x\to 1}\sum_{M=0}^\infty (-x)^M 
=2\lim_{x\to 1} (1+x)^{-1} =1}
from the other terms, which sums to $+3$ as desired. The double pole
of $N^{+-}$ with $N^{+-}$ gets 
a contribution of $+1$ from the first term, $-2$ from the second term,
and 
\eqn\regth{-2(2^2-3^2 +4^2 -5^2 + ...)= -2 -2\sum_{M=0}^\infty M^2 (-1)^M=
-2 -
2\lim_{x\to 1}\sum_{M=0}^\infty M^2 (-x)^M}
from the remaining terms. But by taking derivatives of
$\sum_{M=0}^\infty (-x)^M = (1+x)^{-1}$, one finds
\eqn\formu{\lim_{x\to 1}
\sum_{M=0}^\infty M^2 (-x)^M =\lim_{x\to 1}
( 2 (1+x)^{-3} -3 (1+x)^{-2}
+ (1+x)^{-1}) =0,}
so the $N^{+-}$ double poles sum to $-3$ as desired.

To check the current algebra involving $N^{j-}$, it is
convenient to define
\eqn\Ldef{N^{j-}-\L^{j-} =
-3\p v_{0}^j - v^k_{0} \L^{jk}-
v^j_{0} \L^{+-} +\half  v^k_{0} v^k_{0} w^j_{0}
- v^j_{0} v^k_{0} w^k_{0} +
\half c \s^j_{a\ad} t_{0}^\ad r^a}
$$\equiv a^j_1+ a^j_2 + a^j_3 + a^j_4 + a^j_5 +a^j_6,$$
where $\L^{j-}$ is the second line of $N^{j-}$ in \lj\ and
where 
\eqn\defLJ{\L^{jk}=N^{jk}-v^{[j}_{0}w^{k]}_{0}, \quad
\L^{+-}=N^{+-}-v^{k}_{0}w^{k}_{0}}
are the terms in $N^{jk}$ and $N^{+-}$ which do not involve $v^j_{0}$.
Since $\L^{j-}$ does not involve $v^j_{0}$, one can easily verify
that $N^{j-}$ with $(N^{kl},N^{k+},N^{+-})$ satisfies the current algebra of
\nope.
As usual when constructing Lorentz generators out of light-cone
variables, the most difficult part of the current algebra to check
is that $N^{j-}(y) N^{k-}(z)$ has no singularity.
This will be done by first showing no singularity in 
$(N^{j-}-\L^{j-})(y)
(N^{k-}-\L^{k-})(z)$, then by showing no singularity in
$(N^{j-}-\L^{j-})(y) \L^{k-}(z) + \L^{j-}(y)
(N^{k-}-\L^{k-})(z)$, and finally by showing no singularity in
$\L^{j-}(y)\L^{k-}(z)$.

To show that 
$(N^{j-}-\L^{j-})(y)
(N^{k-}-\L^{k-})(z)$ has no singularity, one can use
$$\L^{jk}(y)\L^{lm}(z)\to {{\d^{l[k} \L^{j]m}(z) -
\d^{m[k} \L^{j]l}(z) }\over {y-z}} - 
{{\d^{m[j} \d^{k]l}} 
\over{(y-z)^2}},
\quad\L^{+-}(y)\L^{+-}(z)\to {5\over{(y-z)^2}},$$
to compute that 
\eqn\compute{
a^j_{2}(y) a^k_{2}(z)\to
{1\over{(y-z)^2}}[\d^{jk}v^l_{0}v^l_{0}-v^j_{0}v^k_{0}]$$
$$+{1\over{(y-z)}}[\d^{jk}v^l_{0}\p v^l_{0}-v^j_{0}\p v^k_{0}
-v^l_{0}\L^{l[j} v^{k]}_{0}-v^l_{0}v^l_{0}\L^{jk}],}
$$ 
a^j_{3}(y) a^k_{3}(z)\to
{5\over{(y-z)^2}}v^j_{0}v^k_{0}+{5\over{(y-z)}}\p
v^j_{0}v^k_{0},$$
$$
a^j_{4}(y) a^k_{4}(z)\to
-{1\over{(y-z)^2}}v^j_{0}v^k_{0}+{1\over{(y-z)}}
[-v^j_{0}\p v^k_{0}+\half
v^l_{0}v^l_{0}v^{[j}_{0}w^{k]}_{0}] ,$$
$$
a^j_{5}(y) a^k_{5}(z)\to
-{11\over{(y-z)^2}}v^j_{0}v^k_{0}-{1\over{(y-z)}}
[v^j_{0}\p v^k_{0}+10v^k_{0}\p v^j_{0}],$$
$$
a^j_{(1}(y) a^k_{4)}(z)\to
-{3\over{(y-z)^2}}\d^{jk}v^l_{0}v^l_{0}-{3\over{(y-z)}}\d^{jk}v^l_{0}\p
v^l_{0},$$
$$
a^j_{(1}(y) a^k_{5)}(z)\to
{6\over{(y-z)^2}}v^j_{0}v^k_{0}+{3\over{(y-z)}}[v^j_{0}\p
v^k_{0}+v^k_{0}\p v^j_{0}],$$
$$
a^j_{(2}(y) a^k_{4)}(z)\to
{1\over{(y-z)}}v^l_{0}v^l_{0}\L^{jk},$$
$$
a^j_{(2}(y) a^k_{5)}(z)\to
{1\over{(y-z)}}v^{[j}_{0}\L^{k]l}v^l_{0},$$
$$
a^j_{(4}(y) a^k_{5)}(z)\to
{2\over{(y-z)^2}}[\d^{jk}v^l_{0}v^l_{0}+v^j_{0}v^k_{0}]$$
$$+{1\over{(y-z)}}[2\d^{jk}v^l_{0}\p v^l_{0}+2v^k_{0}\p v^j_{0}
-\half v^l_{0}v^l_{0}v^{[j}_{0}w^{k]}_{0}],$$
$$
a^j_{(2}(y) a^k_{6)}(z)\to\half
{1\over{(y-z)}}v^{[j}_{0}\s^{k]}_{a\ad}ct^\ad_{0}r^a,$$
$$
a^j_{(3}(y) a^k_{6)}(z)\to -\half
{1\over{(y-z)}}v^{[j}_{0}\s^{k]}_{a\ad}ct^\ad_{0}r^a$$
where 
all functions on the right-hand
side of \compute\ are evaluated at $z$
and $\L^{\mu\nu}$ and $a^j_I$
are defined in \defLJ\ and \Ldef.
Furthermore, one can check that
$$a^j_1(y) a^k_1(z), \quad a^j_6(y) a^k_6(z),\quad a^j_{(1}(y) a^k_{2)}(z),
\quad a^j_{(1}(y) a^k_{3)}(z),\quad a^j_{(2}(y) a^k_{3)}(z),$$
$$
a^j_{(3}(y) a^k_{4)}(z),\quad a^j_{(3}(y) a^k_{5)}(z),\quad a^j_{(1}(y)
a^k_{6)}(z),\quad a^j_{(4}(y) a^k_{6)}(z),\quad a^j_{(5}(y) a^k_{6)}(z) $$ 
have no singularities. One can now easily sum the OPE's of \compute\
to show that 
$(N^{j-}-\L^{j-})(y)
(N^{k-}-\L^{k-})(z)$ has no singularity. 

The next step is to show that 
$(N^{j-}-\L^{j-})(y) \L^{k-}(z) + \L^{j-}(y)
(N^{k-}-\L^{k-})(z)$ has no singularity.
The only contribution comes from
\eqn\onlynext{(a_2^j + a_3^j)(y)\L^{k-}(z) + \L^{j-}(y)
(a_2^k + a_3^k)(z) \to ({1\over{y-z}}+{1\over{z-y}})(\d^{jk} v^l_{0}
\L^{l-} - v_{0}^{(j} \L^{k)-}),}
which has no singularity.
Finally, it will be shown that 
$\L^{j-}(y)\L^{k-}(z)$ has no singularity.

{}From the explicit form of $\L^{j-}$ in the second line of \lj, 
one can check that 
$\L^{j-}(y)\L^{k-}(z)\to (y-z)^{-1} R^{jk}(z)$ where
$R^{jk}$ is cubic in the $(v_M^j,t_M^\ad)$ variables, linear in
the $(w_M^j,u_M^\ad)$ variables,
and does not involve $w_0^j$ or $u_0^\ad$. As will be shown in
section 4, $\widehat Q(\int N^{j-})=0$ where 
$\widehat Q = \int( c\widehat T + s^a\widehat G^a - s^a s^a b)$ and 
$\widehat T$ and $\widehat G^a$ are defined in \ghat.
So $[\int N^{j-},\int N^{k-}] = 
\int R^{jk}$ implies that 
$\widehat Q(\int R^{jk})=0$.
But since $R^{jk}$ does not involve $w_0^j$ or $u_0^\ad$,
$\widehat Q(\int R^{jk})=0$ implies that $R^{jk}=0$.
To prove this, note that 
\eqn\impl{
0=\widehat Q([\int v_0^j,\int R^{kl}]) =
[\int (c v_1^j+s^a \s^j_{a\ad} t^\ad_0),\int R^{kl}] =
[\int c v_1^j,\int R^{kl}]  ,}
$$0=\widehat Q([\int t_0^\ad,\int R^{jk}]) =
[\int (c t_1^\ad +s^a \s^j_{a\ad} v^j_1),\int R^{kl}]
=[\int c t_1^\ad ,\int R^{kl}],$$
which implies that $R^{jk}$ does not involve $w_1^j$ or $u_1^\ad$.
Similarly, one can argue that if $R^{jk}$ is independent of $w_N^j$
and $u_N^\ad$, then it is independent of $w_{N+1}^j$ and $u_{N+1}^\ad$.
So $R^{jk}=0$, which completes the proof that 
$N^{j-}(y)N^{k-}(z)$ has no singularity.

\newsec{Lorentz Invariance of BRST Operator}

In this section, the BRST operator $Q'$ of \brst\ will be shown
to be Lorentz invariant for a certain choice of the coefficients 
$A^{MN j}_{klm}$,
$B^{MN j}_{k\ad\bd}$ and
$C^{MN j}_{\ad\bd k}$ of \lj.
Under commutation with $M^{\mu\nu}$ of \lorentz,
$[s^a, ~\s_j^{a\ad} v^j_{0} s^a + c t_{0}^\ad]$ transform as the
sixteen components of an SO(9,1) spinor and $[-\half(c
+ c v^k_{0} v^k_{0}),~cv^j_{0}, ~ 
-\half (c-
c v^k_{0} v^k_{0})]$ transform as the ten components of an SO(9,1) vector,
so the terms
$[s^a(\g^- d)^a + (\s_j^{a\ad} s^a v^j_{0}+ct_{0}^\ad) (\g^+ d)^\ad]$ and
$[\half c\Pi^- + c v^j_{0} \Pi^j +
\half c v^k_{0} v^k_{0} \Pi^+]$ in $Q'$ are easily seen to be Lorentz
invariant.

Therefore, $Q'$ is Lorentz invariant if $[\int N^{\mu\nu}, \widehat Q]=0$
where $\widehat Q = \int (c\widehat T + s^a \widehat G^a - s^a s^a b)$.
One can easily check that $[\int N^{j+},\widehat Q]=0$ and
$[\int N^{jk},\widehat Q]=0$, so the only remaining question is if one
can define the coefficients in $\L^{j-}$ such that 
$[\int N^{j-},\widehat Q]=0$.
Using the OPE's of \OPE, it is straightforward to compute that
\eqn\straight{[\int(N^{j-}-\L^{j-}),\widehat Q] = \int (c E^j + s^a F^{aj})
\quad{\rm where}}
\eqn\defef{E^j=6\p
v^j_{1}+v^k_{1}
\sum_{M=1}^\infty[(M+1)\d^{jk}v^l_{M}w^l_{M}+v^{[j}_{M}w^{k]}_{M}]}
$$+v^k_{1}\sum_{M=0}^\infty[(M+{3\over
2})\d^{jk}t^\ad_{M}u^\ad_{M}+\half\s^{jk}_{\ad\bd}t^\ad_{M}u^\bd_{M}]
+\half(\s^j\s^k)_{\ad\bd}t^\ad_{0}\sum_{M=0}^\infty[w^k_{M}t^\bd_{M}+
v^k_{M+1}u^\bd_{M}],$$
$$F^{aj}=3\s^j_{a\ad}\p
t^\ad_{0}+\s^k_{a\ad}t^\ad_{0}\sum_{M=1}^\infty[(M+1)\d^{jk}v^l_{M}w^l_{M}
+v^{[j}_{M}w^{k]}_{M}]$$
$$+\s^k_{a\ad}t^\ad_{0}\sum_{M=0}^\infty[(M+{3\over
2})\d^{jk}t^\bd_{M}u^\bd_{M}+
\half\s^{jk}_{\bd\cd}t^\bd_{M}u^\cd_{M}].$$
So one needs to define the coefficients 
$(A^{MN j}_{klm},
B^{MN j}_{k\ad\bd},
C^{MN j}_{\ad\bd k})$ such that 
\eqn\requ{[\widehat Q,\int\L^{j-}] = \int (c E^j + s^a F^{aj}).}

By requiring that both sides of \requ\ coincide for all terms involving
either $v_1^j$ or $t_0^\ad$, one learns that
\eqn\learn{
A^{1M j}_{klm}=
A^{M1 j}_{lkm}=
-{1\over 4}(M+1)(M+2)\d^{jk}\d^{lm}-\half(M+2)\d^{jl}\d^{km}+\half
(M+1)\d^{jm}\d^{kl},}
$$B^{1M j}_{k\ad\bd}=-\half(M+2)^2\d^{jk}\d_{\ad\bd}
-\half(M+2)\s^{jk}_{\ad\bd},$$
$$B^{M0 j}_{k\ad\bd}=-\half(M+3)\d^{jk}\d_{\ad\bd}
-\half(M+1)\s^{jk}_{\ad\bd},$$
$$C^{0M j}_{\ad\bd k}=- C^{M0 j}_{\bd\ad k}=-
{1\over 4}M\d^{jk}\d_{\ad\bd}-{1\over 4}(M+2)\s^{jk}_{\ad\bd}.$$
The only non-trivial check is that the terms 6 $\int c\p v^j_1$ and
$3\int s^a \s^j_{a\ad} \p t^\ad_0$ on the right-hand side of \straight\
are correctly produced by $[\widehat Q,\L^{j-}]$. The first term is
obtained from
\eqn\firstterm{\int c\p v^k (2 \sum_{M=1}^\infty A^{1M j}_{kll}
-\sum_{M=0}^\infty B^{1M j}_{k\ad \ad})}
$$ = (\int c\p v^j)
(2\sum_{M=1}^\infty (-2(M+1)(M+2) -\half (M+2)+\half(M+1))-
\sum_{M=0}^\infty (-4(M+2)^2))$$
$$= (\int c\p v^j)(-\sum_{M=1}^\infty (2M+3)^2 +\sum_{M=0}^\infty (2M+4)^2)
$$
$$= (\int c\p v^j)(\sum_{M=4}^\infty M^2 (-1)^M) =
(\int c\p v^j)(6+\sum_{M=0}^\infty M^2 (-1)^M) = 6\int c\p v^j$$ 
using the result of \formu.
The second term is obtained from
\eqn\secondterm{\int (s^a \p t^\ad_0)\s_k^{a\bd}(\sum_{M=1}^\infty
B^{M0 j}_{k\ad\bd} -2\sum_{M=0}^\infty C^{0M j}_{\ad\bd k})}
$$= 
\int (s^a\s^j_{a\ad}\p t_0^\ad)
(\sum_{M=1}^\infty (-\half (M+3) -{7\over 2}(M+1))
-2 \sum_{M=0}^\infty(-{1\over 4}M -{7\over 4}(M+2)))$$
$$ = 
\int (s^a\s^j_{a\ad}\p t_0^\ad)\sum_{M=3}^\infty (-2M-1)(-1)^M  =
\int (s^a\s^j_{a\ad}\p t_0^\ad)
(3 +\lim_{x\to 1}\sum_{M=0}^\infty (-2M-1)(-x)^M )$$
$$ = 
\int (s^a\s^j_{a\ad}\p t_0^\ad)
(3 +\lim_{x\to 1}(-{2\over{(1+x)^2}} + {1\over{1+x}}))
= 
3\int s^a\s^j_{a\ad}\p t_0^\ad$$
where we used that $(1+x)^{-2} = -\p_x (1+x)^{-1} = 
\sum_{M=0}^\infty M (-x)^{M-1}$. 

Finally, the remaining coefficients in $\L^{j-}$ can be determined
inductively by requiring that all terms in 
$[\widehat Q,\L^{j-}]$ either involve $v_1^j$ or $t_0^\ad$.
This implies that 
\eqn\final{A^{MN j}_{klm} = A^{M (N-1) j}_{klm} + A^{(M-1) N j}_{klm} \quad
{\rm for}~~ N,M >1 ,}
$$B^{MN j}_{k\ad \bd } = B^{M (N-1) j}_{k\ad \bd} + B^{(M-1) N j}_{k\ad \bd}
\quad {\rm for }~~ M>1~~ {\rm and}~~ N>0,$$
$$C^{MN j}_{\ad \bd k } = C^{M (N-1) j}_{\ad \bd k} + C^{(M-1) N j}_{\ad \bd k}
\quad {\rm for }~~ M,N>0.$$

\vskip 15pt
{\bf Acknowledgements:} 
NB would like to thank CNPq grant 300256/94-9, Pronex 66.2002/1998-9
and FAPESP grant
99/12763-0 for partial financial support. OC would like to thank
FONDECYT grant 3000026 for financial support. This research was partially
conducted during the period that NB was employed by the Clay Mathematics
Institute as a CMI Prize Fellow.

\listrefs

\end